\documentclass[twocolumn,prl,showpacs]{revtex4}

\usepackage{graphicx}
\usepackage{rotating}
\usepackage{amsmath}
\usepackage{amsfonts}
\usepackage{amssymb}
\usepackage{enumerate}
\usepackage{longtable}
\usepackage{dcolumn}
\usepackage{bm}

\begin{document}

\newcommand{\be}{\begin{equation}}
\newcommand{\ee}{\end{equation}}
\newcommand{\bn}{\begin{eqnarray}}
\newcommand{\en}{\end{eqnarray}}

\title{Theory of Multi-Band Superconductivity in Iron Pnictides}

\author{M. S. Laad and L. Craco}
\affiliation{Max-Planck-Institut f\"ur Physik komplexer Systeme,
01187 Dresden, Germany}

\date{\rm\today}

\begin{abstract}
The precise nature of unconventional superconductivity in Iron Pnictides is 
presently a hotly debated issue.  Here, using insights from normal state 
electronic structure {\it and} symmetry arguments, we show how an 
unconventional SC emerges from the bad metal ``normal'' state. Short-ranged, 
multi-band spin- and charge correlations generates {\it nodeless} SC in 
the {\it active} planar $d_{xz,yz}$ bands, and an inter-band
{\it proximity} effect induces out-of-plane gap nodes in the {\it passive}
$d_{3z^{2}-r^{2}}$ band. While very good quantitative agreement with various
key observations in the SC state {\it and} reconciliation with NMR and
penetration depth data in the same picture are particularly attractive
features of our proposal, clinching evidence would be an experimental
confirmation of $c$-axis nodes in future work.
\end{abstract}
    

\maketitle

High Temperature Superconductivity (HTSC) in the recently discovered Iron
pnictides (FePn) is the latest surprise among a host of others in $d$- and 
$f$ band materials~\cite{[1]}.  HTSC arises upon doping ($x$) a 
spin-density-wave (SDW) {\it metal}: with $x>x_{c}$, SDW order melts, 
giving way to a HTSC. In the ``normal'' phase, with no SDW/SC order for 
$T>T_{SDW/SC}(x)$, various probes reveal an incoherent metallic state, 
presumably not too far from a Mott insulator~\cite{[2]}.  HTSC in FePn 
arises from an {\it incoherent} normal state, at least in the FeAs-based 
materials, where a quasi-linear-in-$T$ resistivity is seen immediately 
above $T_{c}$: this implies a correlated normal state. Local density
approximation plus dynamical mean-field (LDA+DMFT) calculations 
in the intermediate-to-strong correlation limit~\cite{[3],[4]} indeed find 
very good {\it quantitative} agreement with a range of basic physical 
properties of FePn, attesting to their sizably correlated nature.

{\it If} SC is an instability of this incoherent metal, one expects pronounced
deviations from weak-coupling BCS-like theories in the SC state, as in the
cuprates: this is indeed borne out in experiments.  Specifically,
(i) the pair coherence length, $\xi_{ab}$ is short ($<20~\AA$), the upper
critical field, $H_{c2}$ is high~\cite{[5]}, with small superfluid density,
(ii) the NMR relaxation rate, $T_{1}^{-1}(T)\simeq T^{3}$~\cite{[6]}, and the
penetration depth in the 1111-FePn, $\lambda(T)\simeq T^{n}$ with
$n=2-2.5$~\cite{[7]}, indicating unconventional SC (U-SC).  However, the
122-FePn seem to show conventional $s$-wave-like behavior, though, 
even here, the situation is controversial~\cite{[7]}. Moreover, $\mu$SR 
studies reveal Uemura scaling and $T_{c}$ vs $\epsilon_{F}$ scaling 
characteristic of other, well-known, correlated SC~\cite{[8]}. 
(iii) Optical measurements~\cite{[9]} show a large-scale spectral weight
transfer (SWT) across $T_{c}$ over an energy scale $>O(1.0)$~eV, while 
ARPES data show a sharpening of the low energy quasiparticle kink~\cite{[10]} 
below $T_{c}$. Taken together, as they must be, (i)-(iii) imply a strong 
coupling SC arising from an {\it incoherent} non-FL metal. Given (ii), 
the existence of gap nodes is an open, controversial, issue~\cite{[11]}.
In particular, (i)-(iii) resemble observations in cuprates~\cite{[12]},
implying an U-SC closer to the Bose condensed, rather than the BCS limit~\cite{[10]}. 
How can such a U-SC arise as an instability of an incoherent normal
state~\cite{[3],[4]}?  How may nodes appear in the gap function? Does SC
pairing involve all, or a subset of the Fe-$d$ orbitals?  These issues
are of great import for FePn, and call for a systematic resolution.
Here, motivated by the above, we focus on the 1111-FePn, specifically on 
$LaO_{1-x}FeAsF_{x}$. Guided by
group-theoretical analyses, and by the {\it renormalized} multi-band
structure, we propose a specific, strong correlation based, {\it interband}
pairing mechanism for U-SC as an instability of the incoherent metal found
in earlier LDA+DMFT studies~\cite{[3],[4]}.

LDA+DMFT studies for the 1111-FePn find an incoherent, pseudogapped ``normal''
state, implying {\it blocked} coherent one-electron propagation: only
collective motion of the spin and charge fluid is possible~\cite{[pwa]}.  
This opens
the door to coherent {\it two}-particle propagation, i.e, to two-particle
instabilities, to relieve the finite entropy at lower $T$. At $x=0$, the
electron- and hole-like Fermi surface (FS) sheets are nearly nested in
LDA calculations, and DMFT will not alter this fact, since the
self-energies are {\it local}; this favors the ${\bf q}=(\pi,0)$ SDW
instability, also within LDA+DMFT~\cite{[3]} (in fact, no nesting is needed at
strong coupling~\cite{[2],[3]}), in agreement with neutron
data~\cite{[3],[13]}.  Doping weakens the SDW and facilitates the U-SC 
instability. 

Since coherent one-electron transfer between different orbitals is blocked
in the incoherent metal, we propose, in analogy with the situation encountered
in coupled Luttinger liquids (LL)~\cite{[14]}, that small, residual,
inter-site and inter-orbital interactions mediate two-particle instabilities
in FePn at low $T$.  In principle, various types of instabilities, competing
with U-SC~\cite{[15]}, may be generated, as studied more extensively for
cuprates~\cite{[16]}.  Here, we focus on the U-SC alone.  We further
{\it assume} that pair formation primarily involves the {\it active}
$d_{xz,yz}$ orbitals, and show that, due to the inter-orbital coupling $U'$,
an interband (IB) proximity effect will induce SC on all FS sheets, as in
$Sr_{2}RuO_{4}$~\cite{[17]}. 

Excluding spin-triplet pairing, the general interaction in the cooper channel
is

\be
\nonumber
H_{pair}=\frac{1}{2}\sum_{a,b,k,k'}V_{ab}(k,k')c_{a,k,\uparrow}^{\dag}
c_{b,-k,\downarrow}^{\dag}c_{b,-k',\downarrow}c_{a,k',\uparrow}\;,
\ee
where $a,b=xz,yz$ and the scattering vertex is
$V_{ab}(k,k',\omega)=g^{2}\chi_{ab}(k-k',\omega)$ with
$\chi_{ab}(k-k',\omega)$ being the {\it inter}-orbital susceptibility.
Such a term naturally arises at second order from a one-electron inter-band
term, $t_{ab}\sum_{<i,j>,\sigma}(c_{ia\sigma}^{\dag}c_{jb\sigma}+h.c)$, when
the one-electron spectral function is incoherent, much like in coupled
LL~\cite{[18]}.  This is an important feature of our work: normal state
one-particle incoherence, arising from the Anderson orthogonality
catastrophe~\cite{[3],[4]} due to strong multi-orbital correlations, favors
two-particle coherence. The {\it static}, nearest- and next-nearest neighbor
parts of $V_{ab}(k,k')$ are
$V_{ab}^{(1)}(k,k',0)\simeq\frac{t_{ab}^{2}}{U'+J_{H}}\simeq O(40-50)$~meV and
$V_{ab}^{(2)}\simeq \frac{t_{ab}'^{2}}{U'+J_{H}}\simeq O(15-20)$~meV,
close to the superexchange scale estimated from inelastic neutron scattering
(INS) studies~\cite{[19]}. Notice that the effective interaction contains
coupled inter-orbital charge and spin fluctuations involving nearest (n.n) and
next-nearest neighbor (n.n.n) Fe sites, and is explicitly ${\bf k}$-dependent,
favoring U-SC from the outset. $H_{pair}$ can be decoupled {\it a la}
Gorkov~\cite{[20]} to yield 

\be
\nonumber
H_{pair}^{MF}= \sum_{a,b,k}[\Delta_{ab}(k)c_{k,a,\uparrow}^{\dag}
c_{-k,b,\downarrow}^{\dag}+h.c]\;,
\ee
where $\Delta_{ab}(k)=\frac{1}{2}V_{ab}(k)\langle c_{-k,b,\downarrow} c_{k,a,\uparrow}\rangle$
is the SC gap function.  Consistent with lattice symmetry, and including the
n.n and n.n.n contributions,
$V_{ab}(k,k')=\sum_{l}V_{ab}^{l}\eta_{l}(k)\eta_{l}(k')$, where the
$\eta_{l}(k)$ are irreducible reprersentations of $D_{4h}$ point group. 
So we can expand
$\Delta_{ab}(k)=\sum_{l}\Delta_{ab}^{l}\eta_{l}(k)$~\cite{[20]}.  Explicitly,
$\Delta_{ab}(k)=\Delta_{1}(c_{x}+c_{y})+\Delta_{2}c_{x}c_{y}$, with
$c_{\alpha}=$cos$k_{\alpha}$. The case $\Delta_{2}=0$ ($B_{2g}$ representation
of $D_{4h}$) is favored in the numerical study~\cite{[21]}, however,
$\Delta_{2}\ne 0$ is rigorously {\it required}~\cite{[22]}. 
With well-separated electron- and hole-like FS sheets as in LDA, no
in-plane nodes exist in the SC gap for $\Delta_{2}/\Delta_{1}$ in the chosen
range.  
A $\Delta_{2}\ne 0$ is also
favored by the observation of appreciable geometric frustration in FePn, where
$J_{2}/J_{1}\simeq O(0.7-1.0)$ have been deduced from {\it ab-initio} studies,
and consistent with INS results~\cite{[19]}. 

The ``normal'' state is modelled by a five-band Hubbard model,
$H_{n}=H_{0}+H_{1}$, treated earlier with LDA+DMFT~\cite{[3],[4]}, and reads 
$H_{0}=\sum_{k,\alpha}\epsilon_{k,\alpha}c_{k,\alpha,\sigma}^{\dag}c_{k,\alpha,\sigma}$
while $H_{1}= U\sum_{i,\alpha}n_{i,\alpha,\uparrow}n_{i,\alpha,\downarrow}
+ U'\sum_{i,\alpha\ne \beta}n_{i,\alpha}n_{i,\beta} - J_{H}\sum_{i,\alpha\ne
\beta}{\bf S}_{i,\alpha}\cdot {\bf S}_{i,\beta}$, where $\alpha,\beta=xy,yz,zx,x^{2}-y^{2},3z^{2}-r^{2}$.
In the U-SC phase, we have to solve $H=H_{n}+H_{pair}^{MF}$ within LDA+DMFT. 
Fortunately, the {\it intersite} nature of $H_{pair}$ adds simply a term
bilinear in fermions to $H$.  The LDA+DMFT equations are now readily
extendable to the U-SC regime: the ${\bf G}_{ab}$ and ${\bf \Sigma}_{ab}$
now have normal and anomalous components, and a closed set of DMFT equations,
yielding both $G_{aa},F_{ab}$ selfconsistently, is solved by extending
earlier LDA+DMFT~\cite{[4]} to include an explicit pair-field term. Including
the pair-field, the propagators are 
$G_{aa}(k,\omega)=[\omega-\epsilon_{ka}-\Sigma_{a}(\omega)-\frac{\Delta_{ab}^{2}(k)}{\omega+\epsilon_{kb}+\Sigma_{b}^{\ast}(\omega)}]^{-1}$
and
$F_{ab}(k,\omega)=G_{aa}(k,\omega)\frac{\Delta_{ab}(k)}{\omega+\epsilon_{kb}+\Sigma_{b}^{\ast}(\omega)}$,
where the $\ast$ denotes complex conjugation.
Given {\it intersite} pairing, $F_{ij,ab}(\omega)$ falls off at least as
$\frac{1}{\sqrt{D}}$, (here, $D$ is the lattice dimension) the {\it dynamical}
effects of these non-local pair
fluctuations do not enter the DMFT {\it self-energies}.  Obviously, however,
they do affect the $G_{aa}$s. Our treatment is thus different from that of
Garg {\it et al.}~\cite{[23]}, where the dynamical effects of {\it local}
pair fluctuations must be kept in DMFT. Anomalous contributions like
$\Sigma_{i,j,ab}^{(2)}(t)=-V_{ab}^{2}G_{i,i,aa}(t)F_{i,j,ab}(-t)G_{j,j,bb}(t)$,
etc, to the {\it dynamical} self-energy would thus enter in a cluster-DMFT
approach, but drop out in $D=\infty$.  It then suffices to use the
DMFT with the modified matrix propagators above. Finally, since these
 equations couple {\it all} Fe-$d$-orbitals, the opening up of a SC
gap in the $d_{xz,yz}$ bands could induce secondary gaps in the remaining $d$
orbitals, in a way reminiscent of the inter-band proximity effect in
$Sr_{2}RuO_{4}$~\cite{[17]} (see below). Further, given sizable
$U=4.0$~eV, $U'=2.6$~eV for FePn~\cite{[3],[4]}, large spectral changes
within DMFT-like approaches should accompany the U-SC instability: these
are seen in spectral probes~\cite{[8],[9]}.  Is such a scenario
also consistent with other signatures of a strong coupling, U-SC?

\begin{figure}[thb]
\begin{center}
\includegraphics[width=\columnwidth]{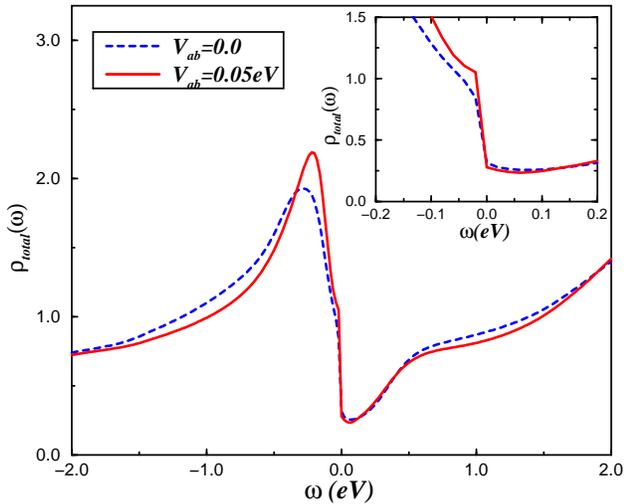}
\end{center}
\caption{(Color online): Evolution of the total one-particle density-of-states
(DOS) across the superconducting transition for doped ($x=0.1$) 
$LaO_{1-x}FeAsF_{x}$, for $U=4.0$~eV, $U'=2.6$~eV, and the effective interaction,
$V_{ab}=$max$(V_{ab}^{(1)},V_{ab}^{(2)})=50$~meV.  The inset shows how 
the low-energy kink, already visible above $T_{c}$ (dashed curve), 
sharpens up below $T_{c}$ (dotted curve), in very good agreement with 
ARPES data~\cite{[10],[24]}.}
\label{fig1}
\end{figure}

Aiming to shine light on these issues, we now describe our results.
Upon convergence of the DMFT equations, the one-electron- {\it and}
pair spectral functions can be read off and used for direct comparison
with observables in the U-SC state. In Fig~\ref{fig1}, we show the
changes induced by U-SC in the {\it total} one-eletron DOS.  Clear
sharpening of the low-energy kink at $\Omega\simeq 20.0$~meV~\cite{[10],[24]},
already somewhat visible in the ``normal'' state~\cite{[25]}, is seen across the
U-SC instability.  Strong normal state incoherence ($\Sigma_{b}^{\ast}(\omega)$
in the equation for $G_{aa}({\bf k},\omega)$) prevents
opening up of a clean SC gap in the DOS, as indeed observed in PES. 
Remarkably, the {\it position} of this kink and the detailed
PES {\it lineshape} up to a binding energy of $-0.5$~eV are both
quantitatively reproduced within our theory.

\begin{figure}[thb]
\begin{center}
\includegraphics[width=\columnwidth]{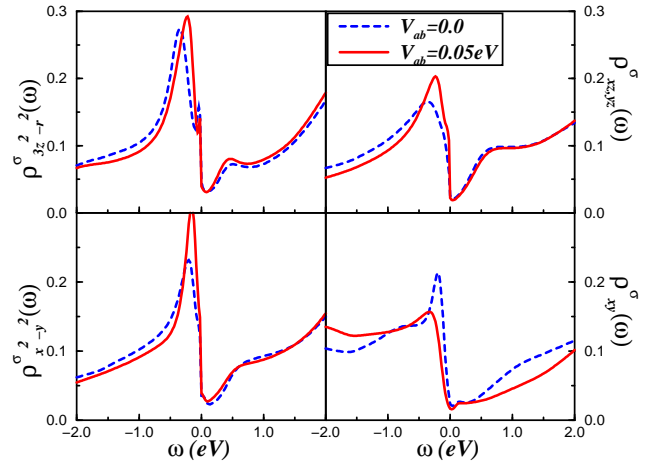}
\end{center}
\caption{(Color online): Orbital resolved evolution of the one-particle DOS
across $T_{c}$, for the same parameter set as in Fig.~\ref{fig1}.  The
orbital-selective gap structure, arising from multi-band correlations, is
manifest.  This is manifested in the orbital-selective sharpening of the 
low-energy kink, in qualitative accord with ARPES~\cite{[10],[24]}. }
\label{fig2}
\end{figure}

More microscopic insight is obtained by investigating the orbital resolved
DOS.  As seen from Fig~\ref{fig2}, only the $d_{xz,yz}$ and $d_{3z^{2}-r^{2}}$
DOS show the sharpening of the kink feature.  Moreover, dominant SWT occurs
{\it from} the $d_{xy}$ band to the $d_{xz,yz,3z^{2}-r^{2}}$ bands, as clearly
seen in Fig.~\ref{fig2}.  The first implies an orbital
{\it selective} coupling of the carrier propagators to multi-orbital, overdamped
and short-ranged, charge- and spin correlations, and fully agrees with
indications from ARPES studies~\cite{[24]}, which show clear evidence thereof.  
Large SWT, over an
energy scale $O(2.0)$~eV, also accompanies the U-SC instability. 
This is also in
qualitative accord with findings in optical studies~\cite{[8]}. Taken together,
these findings imply a strong coupling SC, and fully accord with other
observations in the SC state in FePn, as shown below.

Additional fundamental features of interest are gleaned from observation of the
sharpening of the kink in the $d_{3z^{2}-r^{2}}$ DOS in the U-SC state above. 
This strongly suggests that an IB-proximity driven, out-of-plane SC gap
component should enter on symmetry grounds.  Examination of the
Slater-Koster fit to the LDA shows a sizable $d_{xy}-d_{3z^{2}-r^{2}}$ 
mixing~\cite{[im]}; this will induce a $\Delta'$cos$(k_{z}c)$ component,
reduced by dipolar reduction factors~\cite{[17]}, in the SC gap function.
Strong supportive evidence for this comes from extant dHvA studies on
$LaFePO$~\cite{[26]}.  This indeed shows {\it significant} warping of the
$d_{xy}$ band along the $c$-axis in the {\it non-magnetic} state,
caused precisely by the mixing with the $d_{3z^{2}-r^{2}}$ pocket states.
While correlation effects (LDA+DMFT) result in band shifts and dynamical
SWT, they will leave the LDA FS unchanged, since the self-energies are
purely {\it local}: this observation should hold for the 1111-FePn
in general.  Such interband one-particle mixing between the
$xy$ and $3z^{2}-r^{2}$ bands induces an additional cos$(k_{z}c)$
component in $\Delta_{ab}({\bf k})$, giving possible line nodes in the
SC gap function along $k_{z}$.  This may be theoretically checked by closely 
following earlier analysis for another multi-band U-SC, 
$Sr_{2}RuO_{4}$~\cite{[17]}.

Generation of line nodes along $c$ is {\it not} in conflict with extant ARPES
results, as seen in $(Sr/Ba)_{1-x}K_{x}Fe_{2}As_{2}$~\cite{[10]}. Moreover,
it can naturally explain the NMR and penetration depth data, being capable
of giving the observed power-law-in-$T$ behavior for both the NMR
$T_{1}^{-1}(T)$~\cite{[6]} and $\lambda_{ab}(T)$~\cite{[7]} at low
$T<<T_{c}$.  In contrast to the $s_{\pm}$ gap with disorder
scenario~\cite{[27]}, existence of nodes would imply {\it universal}
power-law dependences in these quantities as a function of $x$, and along
differing FePn members, as recently seen~\cite{[7]}.

We emphasize that whether nodes along $k_{z}$ exist or not is a delicate
matter, and that a careful examination of the details of the band structure
should provide invaluable clues.  For example, dHvA
study~\cite{[26]} on
$LaFePO$ indeed finds sizable $c$-axis corrugation of the $d_{xy}$ band,
indicating sizable IB-proximity, and, interestingly,
$\lambda_{ab}(T)\simeq T^{1.2}$~\cite{[26]}, strongly suggesting line nodes 
in the SC gap.
However, data on $SmFeAsO_{1-x}F_{x}$~\cite{[29]} are consistent with a
{\it smooth} (nodeless) angular variation, 
$\Delta(\phi)=\Delta_{0}(1+\epsilon$cos$4\phi)$, of the in-plane gap, and
$\lambda_{ab}(T)$ follows the well-known $e^{-\Delta(\phi)/kT}$ law. In the FeAs
materials, it is conceivable that the multiband coupling does {\it not} always
induce out-of-plane nodes in the SC gap; this will be sensitive to the
detailed topology of the {\it renormalized} FS. In cases where a
non-exponential $\lambda(T)$ is seen, we suggest that dHvA results should
show significant $c$-axis warping of the $d_{xy}$-FS sheet: this should be 
investigated in more detail. However, the planar $(1+\epsilon$cos$4\phi)$
variation deduced for Sm-based FePn is not inconsistent with
our in-plane form factor, as proposed above. Additionally, recently, INS work 
on $Ni$-doped $BaFe_{2}As_{2}$ clearly shows {\it both} planar and $c$-axis 
variation of $\Delta({\bf k})$; interestingly, precisely
$\Delta(k_{z})\simeq$cos$(k_{z}c)$ was found there~\cite{[11]}. It would be 
very interesting to see if this is also true of the 1111-FePn.  We then 
predict that this should be correlated with the $c$-axis 
warping of planar $d_{xy}$ band in dHvA work.  This would constitute a
non-trivial check of our proposal.

 From LDA+DMFT, we estimate the gap magnitude to be 
$\Delta_{ab}\simeq 4.5$~meV, close to that found in PES~\cite{[ish]}.
Using the carrier Fermi velocity from LDA (note that this is not changed by
LDA+DMFT), $v_{F}\simeq 0.7$~eV.A, the relation $\xi=v_{F}\hbar/\pi\Delta$
gives $\xi\simeq 60-80$A, close to observed values. The upper critical
field is now
estimable as $H_{c2}=\phi_{0}/2\pi\xi^{2}\simeq 50-70$~T, ($\phi_{0}$ is the
usual flux quantum) again consistent with experiment~\cite{[5]}. 
 These are classic 
signatures of a strong coupling SC~\cite{[30]}, and require an 
instability of a strongly correlated {\it incoherent} ``normal'' state, 
as proposed here. 

Our proposal is distinct from other, extant  
ones~\cite{[11],[15],[19],[21],[maz]}. The $s_{\pm}$ proposal~\cite{[15],[maz]}
 is derived from a weak-coupling instability of an {\it itinerant} FL.  Wu
{\it et al.}~\cite{[wu]} derive a similar state from the strong coupling
limit of a reduced two-band model.  The alternative, extended-$s$ wave idea 
gives {\it eight} nodes on the electron-FS sheets~\cite{[11],[19],[21]}.   
Our proposal is a generalization of the $s_{\pm}$ proposal~\cite{[maz],[wu]}. 
 {\it All} earlier proposals, however, cannot access possible $c$-axis nodal 
structure, 
which arises via an IB proximity effect involving $d_{xy}-d_{3z^{2}-r^{2}}$
bands in our work. Moreover, rigorous 
analysis~\cite{[22]} {\it and} the geometrically frustrated superexchanges~\cite{[2],[19]} in FePn require {\it both}, 
$\Delta_{1},\Delta_{2}$ to be finite. This naturally leads to $k$-dependent 
$\Delta_{ab}(k)$, as deduced experimentally~\cite{[11],[29]}. Finally, the 
U-SC derived here arises as an instability of the correlated, incoherent 
metal~\cite{[3],[4]} ``normal'' state seen in doped FePn, and naturally 
explains large spectral changes~\cite{[9]} and orbital selective sharpening 
of the QP kink~\cite{[10]} across $T_{c}$ as manifestations of proximity 
to a Mott insulator~\cite{[2],[3],[4]}.

In conclusion, we have derived the instability of the incoherent ``normal''
state of FePn~\cite{[3],[4]} to an U-SC.  Blocking of inter-band one-particle
{\it coherence} in the normal state clears the way for inter-band
{\it two-particle} coherence to emerge, much like in coupled LLs, giving
a multi-orbital SC with nodes. The latter are shown to arise from a
multiband proximity effect between the $d_{xy}$ and the 
$d_{3z^{2}-r^{2}}$ bands.  Extending earlier LDA+DMFT calculations for the
incoherent metal, we show how good {\it quantitative} agreement
with the sharpening of the low-energy kink in the PES spectrum in the U-SC 
phase, as well as strong spectral weight transfer across the U-SC transition, 
is obtained.  Further, induction of line nodes due to the interband 
proximity effect offers a reconciliation of these attractive features 
with NMR and $\mu$SR data, which suggest existence of line nodes in the 
U-SC gap. Taken together, these strongly support
our proposal of an U-SC with a planar $\Delta_{ab}({\bf k}) \simeq
(c_{x}+c_{y}))+\alpha c_{x}c_{y}$ form
factor; i.e, a SC with co-existent, {\it inter-site} (and inter-orbital) 
$s^{\ast}$ and $s_{xy}$ pair symmetries, and out-of-plane nodes 
[$\Delta_{c}({\bf k}) \simeq c_{z}$], at
least in some of the 1111-FePn family. Our proposal calls for careful 
experimental search for the out-of-plane nodes in the 1111 Iron Pnictides.  



\begin{thebibliography}{28}

\bibitem{[1]} Y. Kamihara {\it et al.},
J. Am. Chem. Soc. 130, 3296 (2008).

\bibitem{[2]} Q. Si and E. Abrahams, Phys. Rev. Lett. {\bf 101},
076401 (2008); ibid. J. Wu {\it et al.}, Phys. Rev. Lett. {\bf 101}, 126401
(2008); G. Baskaran, J. Phys. Soc. Jpn. {\bf 77}, 113713 (2008);
 Q. Si {\it et al.}, arXiv:0901.4112. 

\bibitem{[3]} K. Haule, {\it et al.}, Phys. Rev. Lett.
{\bf 100}, 226402 (2008).

\bibitem{[4]} L.Craco {\it et al.},
Phys. Rev. B {\bf 78}, 134511 (2008); ibid
M. S. Laad {\it et al.},
Phys. Rev. B. {\bf 79}, 024515 (2009).

\bibitem{[5]} J. Jaroszynski {\it et al.}, 
Phys. Rev. B {\bf 78}, 174523 (2008).  

\bibitem{[6]} T. Imai {\it et al.},
J. Phys. Soc. Jpn. {\bf 77}, 47 (2008).

\bibitem{[7]} H. Luetkens {\it et al.},
Phys. Rev. Lett. {\bf 101}, 097009 (2008); J.D. Fletcher {\it et al.},
arXiv:0812.3858.

\bibitem{[8]} Y.J. Uemura, arXiv:0811.1546. 

\bibitem{[9]} A.V. Boris {\it et al.},
Phys. Rev. Lett. {\bf 102}, 027001 (2009);
ibid S.I. Mirzaei {\it et al.}, arXiv:0806.2303.

\bibitem{[10]} L. Wray {\it et al.},
Phys. Rev. B {\bf 78}, 184508 (2008).

\bibitem{[11]} S. Chi {\it et al.}, arXiv:0812.1354.  In S. Graser 
{\it et al.}, arXiv:0812.0343, and V. Mishra {\it et al.}, 0901.2653, 
the pair wave function is argued to have ex-$s$ form, and that disorder 
lifts the nodal structure.  The opposite is argued in 
D. Parker {\it et al.}, Phys. Rev. B {\bf 78}, 134524 (2008). 

\bibitem{[12]} P.A. Lee {\it et al.},
Rev. Mod. Phys. {\bf 78}, 17 (2006).

\bibitem{[pwa]} P.W. Anderson, {\it The Theory of Superconductivity in the 
High-$T_{c}$ Cuprates}, Princeton Univ. Press (1997).

\bibitem{[13]} C. de la Cruz {\it et al.},                                  
Nature {\bf 453}, 899 (2008).

\bibitem{[14]} A. Nersesyan {\it et al.}, in 
{\it Bosonization and Strongly Correlated Electronic Systems}, 
Cambridge University Press (1998). 

\bibitem{[15]} A. Chubukov {\it et al.}, Phys. Rev. B {\bf 78}, 
134512 (2008), where this issue is considered in the weak coupling 
limit.  

\bibitem{[16]} C. Xu {\it et al.},
Phys. Rev. B {\bf 78}, 020501 (2008); ibid C. M. Varma, 
Phys. Rev. B {\bf 73}, 155113 (2006); H. Yamase and 
W. Metzner, Phys. Rev. B {\bf 75}, 155117 (2007).

\bibitem{[17]} M.E. Zhitomirsky and T.M. Rice, Phys. Rev.
Lett. {\bf 87}, 057001 (2001).

\bibitem{[18]} D. V. Khveshchenko, 
Phys. Rev. B {\bf 50}, 380 (1994).

\bibitem{[19]} M.M. Parish {\it et al.},
Phys. Rev. B {\bf 78}, 144514 (2008); ibid. T. Yildirim, 
Phys. Rev. Lett. {\bf 101}, 057010 (2008).

\bibitem{[20]} R. Fehrenbacher and M. R. Norman,
Phys. Rev. Lett. {\bf 74}, 3884 (1995).

\bibitem{[21]} M. Daghofer {\it et al.},
Phys. Rev. Lett. {\bf 101}, 237004 (2008);
A. Moreo {\it et al.},
arXiv:0901.3544.

\bibitem{[22]} Wen-Long You {\it et al.}, arXiv:0807.1493.

\bibitem{[23]} A. Garg {\it et al.},
Phys. Rev. B {\bf 72}, 024517 (2005).

\bibitem{[24]} P. Richard {\it et al.}, 
Phys. Rev. Lett. {\bf 102}, 047003 (2009).

\bibitem{[25]} H. W. Ou {\it et al.}, 
Sol. St. Commun. {\bf 148}, 504 (2008).

\bibitem{[im]} K. Nakamura {\it et al.}, J. Phys. Soc. Jpn. {\bf 77}, 093711 
(2008).

\bibitem{[26]} A.I. Coldea {\it et al.},
Phys. Rev. Lett. {\bf 101}, 216402 (2008); ibid. A. Carrington {\it et al.}, 
arXiv:0901.3976.

\bibitem{[27]} D. Parker {\it et al.}, Phys. 
Rev. B {\bf 78}, 134524 (2008).

\bibitem{[maz]} I.I. Mazin {\it et al.},
Phys. Rev. Lett. {\bf 101}, 057003 (2008); I.I. Mazin and J. Schmalian,
arXiv:0901.4790.

\bibitem{[wu]}  J. Wu and P. Phillips, arXiv:0901.3538.

\bibitem{[29]} L. Malone {\it et al.}, arXiv:0806.3908.

\bibitem{[ish]} Y. Ishida {\it et al.}, Phys. Rev. B {\bf 79}, 060503 (2009).
 
\bibitem{[30]} J. Bauer {\it et al.}, arXiv:0901.1760.

\end{thebibliography}
\end{document}